\documentclass[a4, table]{article}
\usepackage{spconf,amsmath,graphicx}
\usepackage{amssymb}
\usepackage{times}
\usepackage{amsfonts}
\usepackage{pdfpages}
\usepackage{subcaption}
\usepackage[export]{adjustbox}
\usepackage{array}
\usepackage{booktabs}
\usepackage{hyperref}
\usepackage[most]{tcolorbox}
\usepackage{makecell}

\usepackage{fancyhdr}

\usepackage[dvipsnames]{xcolor}
\usepackage{highlight}
\newcommand{\g}[1]{\gradientcelld{#1}{0}{0.4}{1}{red}{white}{CornflowerBlue}{70}}
\newcommand{\grmse}[1]{\gradientcelld{#1}{0}{30}{30}{CornflowerBlue}{white}{white}{70}}

\newcommand{\esp}[1]{\mathbb{E}\left[#1\right]}

\def\tf{{t, f}}
\newcommand{\figref}[1]{Fig.~\ref{#1}}

\newcommand{\ie}{\emph{i.e.}}
\newcommand{\eg}{\emph{e.g.}}

\newcommand\blfootnote[1]{%
  \begingroup
  \renewcommand\thefootnote{}\footnote{#1}%
  \addtocounter{footnote}{-1}%
  \endgroup
}

\definecolor{block-gray}{gray}{0.85}
\newtcolorbox{proof}{colback=block-gray,grow to right by=-1mm,grow to left by=-1mm,
boxrule=0pt,boxsep=0pt,breakable}

\title{QASTAnet: A DNN-based Quality Metric for Spatial Audio}
\name{Adrien Llave, Emma Granier, Grégory Pallone}
\address{Orange Research, France}

\begin{document}
\ninept
\maketitle
\begin{abstract}
    
In the development of spatial audio technologies, reliable and shared methods for evaluating audio quality are essential.
Listening tests are currently the standard but remain costly in terms of time and resources.
Several models predicting subjective scores have been proposed, but they do not generalize well to real-world signals.
In this paper, we propose QASTAnet (Quality Assessment for SpaTial Audio network), a new metric based on a deep neural network, specialized on spatial audio (ambisonics and binaural).
As training data is scarce, we aim for the model to be trainable with a small amount of data.
To do so, we propose to rely on expert modeling of the low-level auditory system and use a neural network to model the high-level cognitive function of the quality judgement.
We compare its performance to two reference metrics on a wide range of content types (speech, music, ambiance, anechoic, reverberated) and focusing on codec artifacts.
Results demonstrate that QASTAnet overcomes the aforementioned limitations of the existing methods.
The strong correlation between the proposed metric prediction and subjective scores makes it a good candidate for comparing codecs in their development.
We provide a fully differentiable open-source implementation in Python/Pytorch (\url{https://github.com/Orange-OpenSource/QASTAnet}).
This way, the metric can be used as a training objective for the development of spatial audio processing and can take advantage of GPU acceleration.
\end{abstract}

\blfootnote{© 2025 IEEE. Personal use of this material is permitted. Permission from IEEE must be obtained for all other uses, in any current or future media, including reprinting/republishing this material for advertising or promotional purposes, creating new collective works, for resale or redistribution to servers or lists, or reuse of any copyrighted component of this work in other works.}
\section{Introduction}
\label{sec:intro}

In the design of spatial audio processing, \eg, audio codecs, it is crucial to rely on a proven protocol to measure the quality of encoded/decoded signals.
The standard in the community is the listening test conducted on a group of listeners.
The two most common are the Degradation Category Rating (DCR) test with naïve subjects and the Multi-Stimuli with Hidden Reference and Anchors (MUSHRA) test~\cite{noauthor_itu-r_2015} with expert subjects.

Listening tests are particularly costly and time-consuming.
For this reason, many studies have attempted to develop algorithms to predict the average results of subjective tests.
In mono, POLQA (ITU-T Rec. P.863)~\cite{noauthor_itu-t_2018} effectively predicts the degradation of quality due to codecs on speech~\cite{muller_speech_2024}.
However, for spatialized signals such as (higher-order) ambisonic or binaural, although metrics including the spatial dimension have been proposed~\cite{george_feature_2006, choi_objective_2008, dewhirst_qestral_2008, delgado_objective_2019, manocha_spatialization_2023}, none have yet been widely adopted in subsequent works, where subjective tests remain preferred~\cite{hold_perceptually-motivated_2024, vasilache_metadata-assisted_2025}.
It is worth mentioning that a detailed review of the topic was conducted recently~\cite{rafaely_loss_2025}.
In the present study, we focus particularly on the ambisonic format~\cite{daniel_representation_2001}.
This representation format is useful for immersive content, particularly in the field of augmented/virtual reality, as it allows for easy rotations of the sound scene.
In the following, we focus our empirical comparison on methods with publicly available code. 
Nonetheless, we discuss related methods without code for context.

The Ambiqual metric~\cite{narbutt_ambiqual_2018, narbutt_ambiqual_2020} is designed to predict MUSHRA scores on codec degradations for ambisonic signals rendered on headphones based on two criteria: listening quality (LQ) and localization accuracy (LA).
Its specificity is to be independent of the rendering method, \eg, speakers setup, binauralization, by directly comparing reference and degraded ambisonic signals.
The metric evaluation uses only spatialized signals comprised of plane waves, including speech, music, and bursts of pink noise.
The signals do not contain any spatially diffuse components, such as those typically introduced by a natural content or the convolution with spatial room impulse responses (SRIR).
Since the metric relies on the comparison of phaseograms, it is expected to struggle in accurately predicting scores for signals with realistic reverberation.
Indeed, parametric codecs~\cite{weckbecker_ambisonics_2025} which produce uncorrelated signals often do not preserve their phase.

The eMoBi-Q metric~\cite{eurich_computationally_2024} was introduced to evaluate processing applied to binaural signals, particularly within augmented listening contexts.
The method relies on a sophisticated model of the auditory system for low-level features extraction.
However, the modeling of the quality judgement from those features is not straightforward.
The method encompasses a simple yet effective handcrafted function that demonstrates strong predictive performance for the signals and degradations under consideration.
This study limits the evaluation to speech signals and does not test the predictive capability of the metric on codec artifacts.
In particular, the authors warn that eMoBi-Q cannot take into account non-linear artifacts, such as those introduced by codecs.

The SPAUQ metric~\cite{watcharasupat_quantifying_2024} aims to assess the spatial and non-spatial impairment of any multichannel signals in an \emph{SNR-fashion} in line with BSSEval~\cite{vincent_performance_2006}.
It is not designed to predict subjective test, so we do not take it into account in our study.

The Generative Machine Listener (GML)~\cite{jiang_generative_2023} is a model based on a convolutional neural network with 15 million trainable parameters.
To prevent overfitting, training required 67k examples and a data augmentation technique assuming that the MUSHRA score of a mix of two signals is the same as the mixture of their corresponding scores.
No implementation is available, and the large amount of data required to train the model makes it difficult to reproduce.

\begin{figure*}[t]
    \includegraphics[trim=100 50 50 150, clip, width=\textwidth]{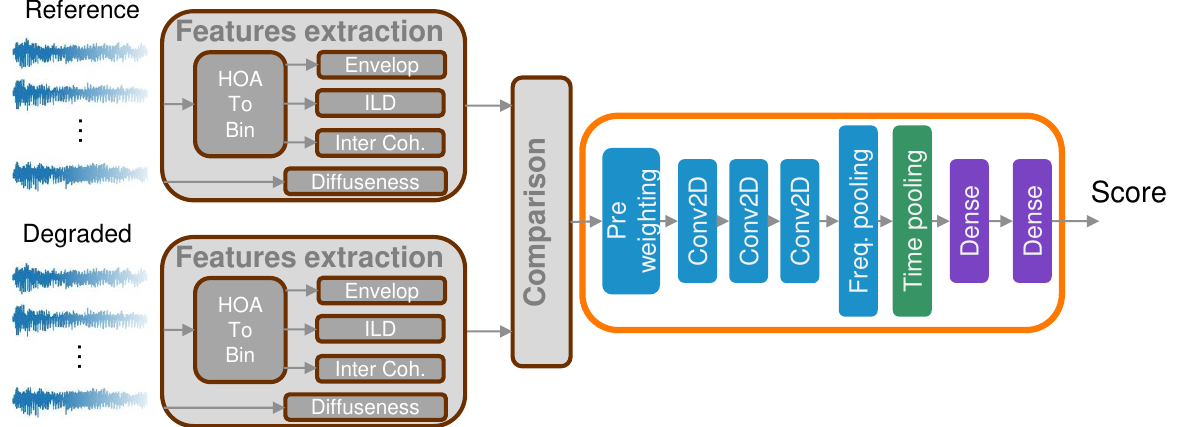}
    \caption{\label{fig:architecture}Schematic representation of the QASTAnet metric architecture.}
\end{figure*}

In this work, we propose QASTAnet, a metric for evaluating overall audio quality based on a neural network for spatial audio.
Given the limited availability of training data, we aim for the model to be trainable with a small amount of data.
We present the architecture based on expert features extraction and a small Deep Neural Network (DNN).
This way, our metric fills the gap between a pure knowledge based approach (\eg, eMoBi-Q) and a pure end-to-end deep learning approach (\eg, GML).
Indeed, we assume that for approximating the cognitive function between low-level auditory features and quality judgement, a deep learning approach can be more effective than an handcrafted one as in eMoBi-Q.
To train the DNN, we created a dataset based on a MUSHRA test.
Finally, we compare QASTAnet to two metrics whose code is available: Ambiqual and eMoBi-Q.
The evaluation covers a wide range of signals (speech, music, and ambiance) and realistic conditions (synthetic mixing with and without diffuse SRIR, native recording using ambisonic microphones).
The study focuses on 3\textsuperscript{rd}-order ambisonic signals and codec-induced degradations.

\section{Proposed method}

\subsection{Architecture}

\figref{fig:architecture} presents an overview of the proposed model architecture.
To limit the number of trainable parameters of the model, we choose to use expert features upstream of the neural network.
We mainly rely on the time-frequency (T-F) features from eMoBi-Q~\cite{eurich_computationally_2024}: the monaural envelope, interaural level difference (ILD), and interaural coherence.
These three features require prior binauralization of ambisonic signals.
Since binauralization is a very specific encoding of spatial information with a significant dimensionality reduction, \eg, 16 to 2 for 3\textsuperscript{rd}-order ambisonics, we assume that adding a feature in the ambisonic domain can improve performance.
So, based on the complex T-F representation of 1\textsuperscript{st}-order ambisonic signals with components $w_\tf, x_\tf, y_\tf$, and $z_\tf$ with $t$ and $f$ representing the temporal and frequency indices, we propose to estimate the diffuseness~\cite{pulkki_parametric_2017} defined as:
\begin{equation}
    \Psi_\tf = 1-\frac{||\esp{\mathbf{I}_\tf}||}{\esp{e_\tf}},
\end{equation}
where $\mathbf{I}_\tf=\text{Re}\lbrace w_\tf^* . \left[ x_\tf, y_\tf, z_\tf \right]^T\rbrace$ is the intensity vector and $e_\tf=|w_\tf|^2+\frac{|x_\tf|^2+|y_\tf|^2+|z_\tf|^2}{2}$ is the energy of the ambisonic signal.
In practice, the expectation $\esp{.}$ is approximated by a first-order recursive filter described in \cite{pulkki_parametric_2017}.
The eMoBi-Q metric uses a temporal resolution of 400~ms without overlap.
We hypothesize that, in order to capture coding artifacts and to analyze a complex scene, a lower temporal resolution is necessary.
So we choose to use a temporal resolution of 40~ms for all features.

The reference/degraded features are then compared using a quadratic difference.
Next, a learned weighting, called \emph{pre-weighting}, is applied to each feature for each frequency band.
Then, a block of 3 point-wise T-F convolutional layers extracts a higher-level representation.
Each convolutional layer includes a LeakyReLU activation, and the number of filters in each layer is 16, 16, and 6, respectively.
The resulting representation is averaged over the frequency dimension with a learned weighting.
Then, a softmax-weighted average pooling layer~\cite{mcfee_adaptive_2018} acts to reduce the temporal dimension.
A dense layer followed by a LeakyReLU projects the resulting 6-dimensional vector to 16 dimensions.
Finally, a dense layer followed by a sigmoid produces the estimated MUSHRA score scaled between 0 and 1.
In total, the network has 730 trainable parameters.
The predicted score may vary depending on the set of binauralization filters (HOA2bin) used for extracting binaural features.
To limit this effect, we improved results by averaging the metric's scores using multiple instances of the same model with up to 20 different sets of HOA2bin filters, \ie\ 20 different \emph{heads}.
These filters are obtained thanks to the MagLS method~\cite{schorkhuber_binaural_2018} from 11 publicly available \cite{carpentier_measurement_2014} and 9 internal sets of Head-Related Transfer Functions (HRTFs) measured in anechoic room.

\subsection{Dataset}
Since we did not identify an open source database providing both HOA content and associated scores (most of them being mono or even stereo), we set up a MUSHRA test to construct the dataset.
It is divided into 6 sessions (1 for testing, 1 for validation, 4 for training) comprising 13 stimuli and 7 degradations, for a total of 546 examples, 364 of which are for training.
The stimuli are selected to be as varied as possible, including speech (one or multiple speakers), music (orchestra, small ensemble), and ambiances, \eg, applause, party, park.
The spatialization techniques are also diverse: ideal plane wave encoding, convolution by SRIR of spherical microphone arrays (SMA), and native recording with SMA.
Each stimulus has a duration of approximately 10~s and a global level of -30~LUFS~\cite{noauthor_ebu_2023} (after binauralization), sampled at 48~kHz and truncated to 3\textsuperscript{rd}-order ambisonic if necessary.

For the conditions under test (CuT), we consider the artifacts introduced by the IVAS codec in SBA mode~\cite{noauthor_codec_2024, weckbecker_ambisonics_2025} (at bitrates 32, 64, 128, and 256~kbps), EVS applied independently to each ambisonic channel~\cite{dietz_overview_2015} (at bitrates 16$\times$16.4~kbps and 16$\times$32~kbps), and a low-pass filtered anchor at 3.5~kHz.

Two groups of participants took part in the MUSHRA test, one comprising 6 subjects who completed all 6 sessions, and the other consisting of 13 subjects, who only completed session~1.
This allows us to evaluate whether the performance of the metric generalizes to a group of participants whose results were not used for training.

The listening test took place in a listening studio of 40~m$^2$ with a ceiling height of 4~m.
Both the background noise below the NR15 curve and the RT$_{60}$ of 0.29~s comply with the ITU-R~BS.1116~\cite{noauthor_itu-r_2015-1}.
The signals are played through a dome of 29 Amadeus PMX5 loudspeakers arranged with a radius of 2.7~m.
The loudspeakers are evenly distributed on horizontal rings at 5 different elevations: 4 at -22°, 12 at 0°, 8 at 25°, 4 at 60°, and 1 at 90°.
For all rings, one speaker is placed at azimuth 0°.
Ambisonic decoding is performed using the AllRADecoder plug-in from the IEM suite\footnote{\url{https://plugins.iem.at/}} with the maxrE method~\cite{zotter_all-round_2012}.
The system is calibrated to ensure that 65~dB(A) corresponds to -30~LUFS for spatialized pink noise in the frontal direction at the listening position, and listeners can freely adjust the level by $\pm$~4dB.

\subsection{Training and loss functions}

Based on the results from the four MUSHRA test sessions dedicated to training, the model is trained in a supervised manner using the Adam optimizer with a learning rate of $0.003$ and a batch size of 32.
The Mean Square Error (MSE) between the model's output and the mean MUSHRA score, scaled between 0 and 1, is used as the cost function.
Training is halted when the Pearson correlation coefficient $r$ calculated on the validation set has not improved over the last 15 epochs.
On average, 2.5k training steps are performed, equivalent to 220 epochs.
In the end, we retain the model that maximized $r$ on the validation set across all epochs.
Hidden reference data are excluded from the training to avoid wasting network capacity for a useless task.

\section{Evaluation}

\subsection{Evaluation criteria}  

To evaluate the accuracy of a metric, we compare its predictions to MUSHRA scores averaged across listeners according to the following criteria:
(i) the Pearson correlation coefficient $r$, measuring the linear correlation between prediction and ground truth;
(ii) the Spearman correlation coefficient $\rho$, measuring the rank correlation between predictions and ground truth;
(iii) the root mean square error (RMSE);
(iv) the \emph{epsilon-insensitive} RMSE (RMSE*)~\cite{noauthor_itu-t_2020}, which takes into account the mean estimate uncertainty:
\begin{equation}
    \text{rmse}^\star = \sqrt{\sum\limits_i \text{max}\left(0, |s_i - \hat{s}_i|-\text{ci95}_i\right)^2},
\end{equation}
where $s_i$ is the empirical MUSHRA mean score for the $i^\text{th}$ example, $\hat{s}_i$ the corresponding model mean estimate and ci95, the empirical mean confidence interval at 95~\%.

\subsection{Baselines}  

We consider metrics that predict overall audio quality from spatialized sounds for which the code is available.
We retain the Ambiqual~\cite{narbutt_ambiqual_2020} and eMoBi-Q~\cite{eurich_computationally_2024} metrics.
Ambiqual directly handles 3\textsuperscript{rd}-order ambisonic signals.
The RMSE requires comparing two distributions on the same scale, but Ambiqual LA and LQ are not natively between 0 and 100.
Following results figures in~\cite{narbutt_ambiqual_2020}, we apply a linear scaling by $\frac{100}{0.3}$ for LA and 100 for LQ.
Since eMoBi-Q processes binaural signals, we must binauralize the ambisonic signals beforehand.
To do this, we use the aforementioned HOA2bin filters.

\subsection{Results}  

\begin{figure}[t]
    \includegraphics[width=.5\textwidth]{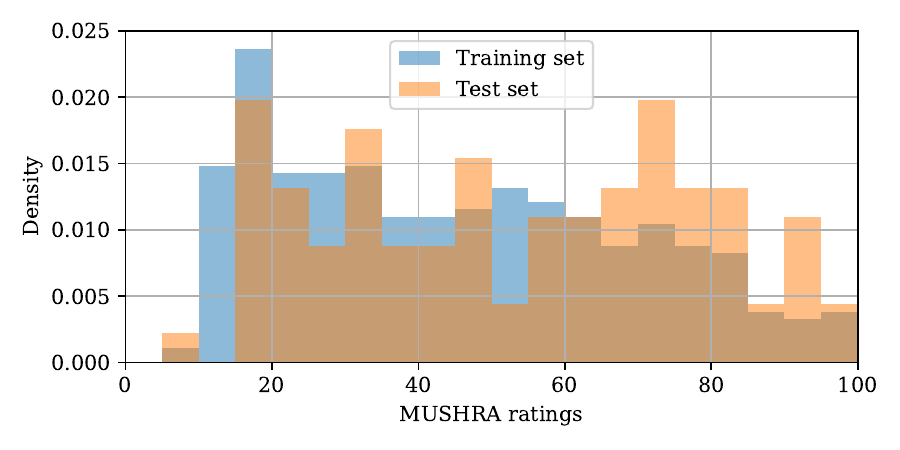}
    \caption{\label{fig:dataset_labels_hist}Histograms of MUSHRA ratings for the training and test sets. The hidden reference ratings are excluded.}
\end{figure}

\begin{figure*}[t]
    \centering
    \includegraphics[width=.4\textwidth, trim={0.5cm 0.5cm 0.5cm 0.5cm}, clip]{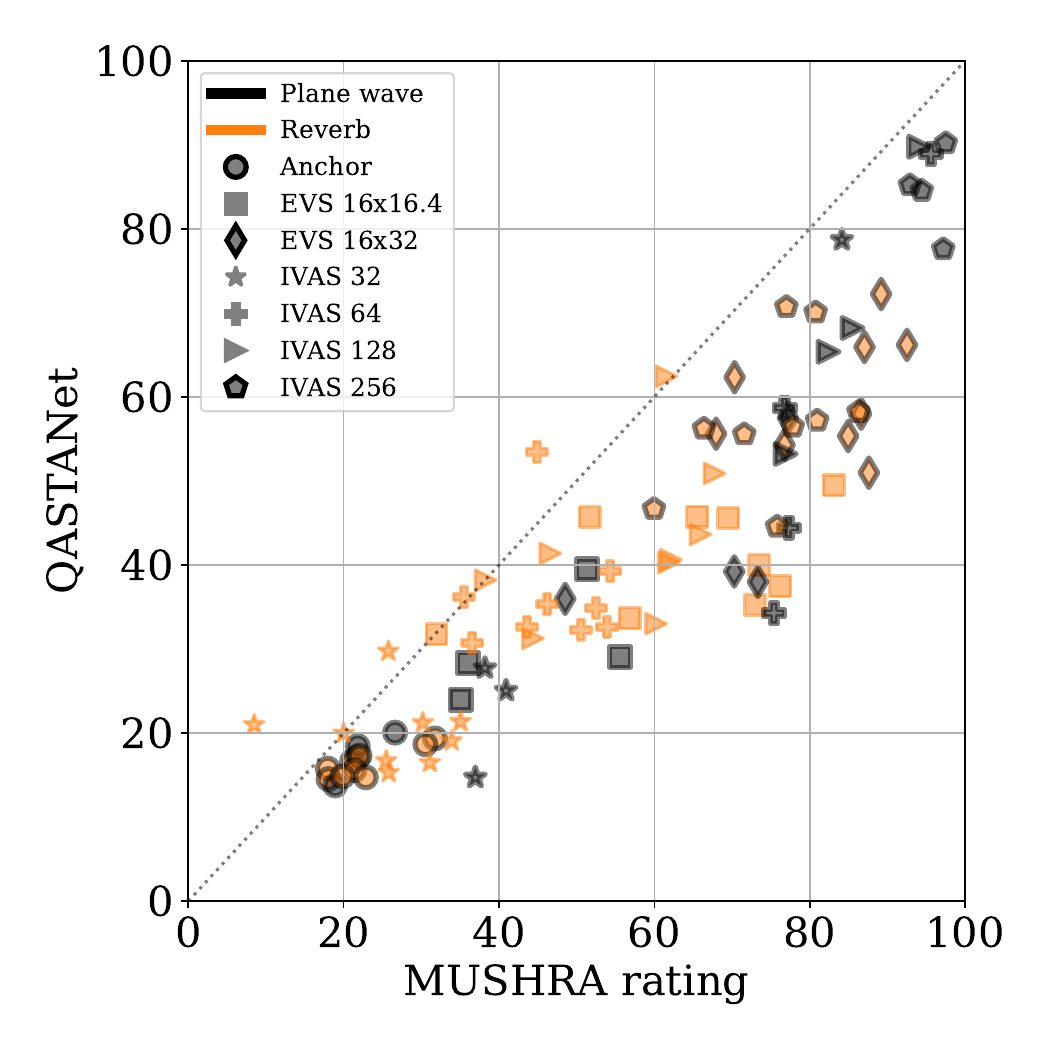}
    \includegraphics[width=.4\textwidth, trim={0.5cm 0.5cm 0.5cm 0.5cm}, clip]{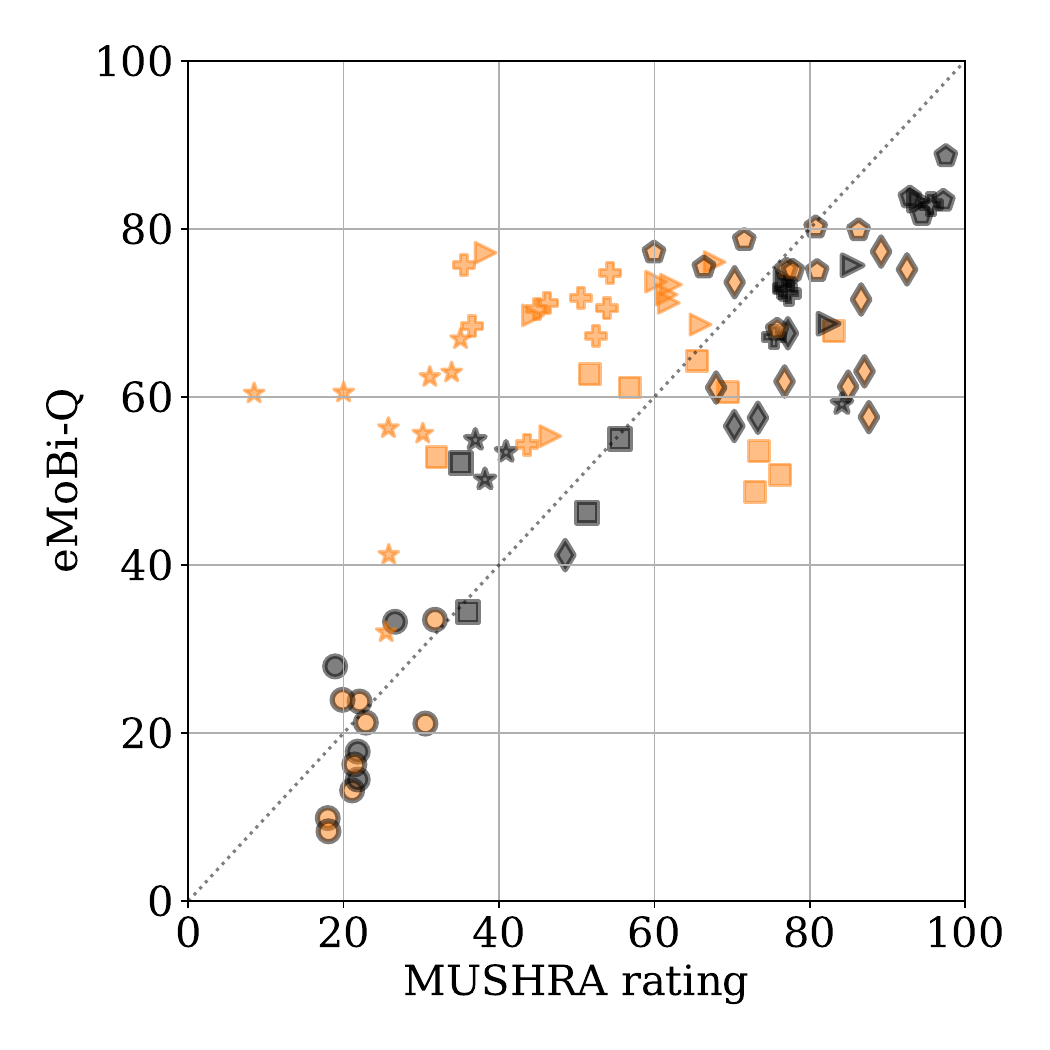}
    \caption{\label{fig:metrics_vs_mushra}QASTAnet (left) and eMoBi-Q (right) predictions vs. MUSHRA rating for each combination of stimulus and CuT. The evaluation does not consider the hidden reference ratings. The orange and black markers distinguish the signals including spatial reverberation from those generated with an ideal plane wave encoding (anechoic), respectively.}
\end{figure*}

\begin{table*}[t]
    \small
    \centering
    \caption{\label{tab:results_corr}Evaluation criteria between objective metrics predictions and subjective rating. We show the results considering three subsets of CuT or stimulus, see text for further details. The best of each column is bolded. The color intensity of cells is mapped from 0.4 to 1 for the correlation coefficients and from 30 to 0 for the RMSE.}
    
    \begin{tabular}{l | c c c | c c c | c c c | c c c }
    \hline\hline
    Metric & \multicolumn{3}{c}{Pearson $\uparrow$}& \multicolumn{3}{c}{Spearman $\uparrow$}& \multicolumn{3}{c}{rmse $\downarrow$}& \multicolumn{3}{c}{rmse* $\downarrow$}\\
    & \scriptsize all & \scriptsize codecs & \scriptsize spat. rev. & \scriptsize all & \scriptsize codecs & \scriptsize spat. rev. & \scriptsize all & \scriptsize codecs & \scriptsize spat. rev. & \scriptsize all & \scriptsize codecs & \scriptsize spat. rev. \\
    \hline
    Ambiqual LA & \g{0.61} & \g{0.77} & \g{0.58} & \g{0.58} & \g{0.82} & \g{0.56} & \grmse{28.8} & \grmse{30.2} & \grmse{29.2} & \grmse{25.4} & \grmse{26.8} & \grmse{25.7} \\
    Ambiqual LQ & \g{0.51} & \g{0.48} & \g{0.40} & \g{0.50} & \g{0.48} & \g{0.37} & \grmse{22.1} & \grmse{22.1} & \grmse{21.6} & \grmse{19.3} & \grmse{19.4} & \grmse{18.7} \\
    eMoBi-Q & \g{0.72} & \g{0.55} & \g{0.63} & \g{0.71} & \g{0.56} & \g{0.59} & \grmse{17.3} & \grmse{18.5} & \grmse{19.4} & \grmse{14.4} & \grmse{15.5} & \grmse{16.4} \\
    \hline
    QASTANet & \textbf{\g{0.90}} & \textbf{\g{0.86}} & \textbf{\g{0.89}} & \textbf{\g{0.92}} & \textbf{\g{0.88}} & \textbf{\g{0.89}} & \grmse{18.4} & \grmse{19.7} & \grmse{18.4} & \grmse{15.3} & \grmse{16.5} & \grmse{15.2} \\
    \quad wo/ pre-weighting & \g{0.89} & \g{0.85} & \g{0.86} & \g{0.91} & \g{0.87} & \g{0.87} & \textbf{\grmse{16.0}} & \textbf{\grmse{17.1}} & \textbf{\grmse{16.2}} & \textbf{\grmse{13.0}} & \textbf{\grmse{14.0}} & \textbf{\grmse{13.3}} \\
    \quad wo/ diffuseness feat. & \g{0.88} & \g{0.84} & \g{0.87} & \g{0.90} & \g{0.86} & \g{0.86} & \grmse{19.4} & \grmse{20.8} & \grmse{19.5} & \grmse{16.4} & \grmse{17.7} & \grmse{16.4} \\
    \quad frame length 400~ms & \g{0.85} & \g{0.79} & \g{0.81} & \g{0.88} & \g{0.82} & \g{0.84} & \grmse{19.4} & \grmse{20.9} & \grmse{19.4} & \grmse{16.6} & \grmse{17.9} & \grmse{16.6} \\
    \hline\hline
    \end{tabular}

\end{table*}

Tab.~\ref{tab:results_corr} presents the evaluation criteria for three subsets of stimuli and CuT combinations: \emph{all} includes all signals, \emph{codecs} excludes low-pass filtered anchor signals, and \emph{spat. rev.} includes only signals recorded with an SMA or artificially spatialized thanks to SRIR.
\figref{fig:metrics_vs_mushra} shows the detailed results for the QASTAnet and eMoBi-Q metrics.

\subsubsection{Metrics comparison}
First, we observe that Ambiqual LA has a good correlation ($r$=0.77) with codec degradations but fails to predict the quality drop on the anchor (0.61) as it only concerns timbre modifications.
We also note that Ambiqual LQ mildly correlates the subjectives ratings on signals containing spatially realistic reverberation (0.40), dropping down by 0.11 with the exclusion of anechoic signals.
We expected this phenomenon as Ambiqual relies on phaseograms comparison in the T-F domain, as explained earlier.
To our knowledge, previous works did not investigate this aspect.
In \figref{fig:metrics_vs_mushra}, one can see that eMoBi-Q performs particularly well on anechoic signals and accurately predicts the anchor degradation.
However, the correlation drops down to 0.63 for the \emph{spat. rev.} subset because it overestimates IVAS at the lowest bitrates (32 and 64~kbps) and underestimates the multi-mono EVS degradations.

Next, QASTAnet outperforms both baselines in terms of correlation coefficients across all three signal subsets.
However, QASTAnet exhibits a higher RMSE than eMoBi-Q.
The \figref{fig:metrics_vs_mushra} indicates that the mismatch between both metric families can be attributed by a slope bias between QASTAnet and the test set ratings, leading to an underscoring effect.
Meanwhile, the QASTAnet scores dispersion appears smaller than that of eMoBi-Q.
This explains the higher correlation as $r$ is invariant to an affine transformation and $\rho$ is invariant to any monotonic transformation.
To explain this bias, we have two hypotheses. 
First, the six participants involved in collecting the training data graded more strictly than those involved exclusively in the test data.
In particular, the training group tends to assign lower ratings to signals processed by EVS.
To limit a possible bias propagation due to the small number of participants for the training set, we tested a mitigation strategy based on training with the RMSE* criterion, but no bias reduction was observed.
Second, the training labels are not uniformly distributed with a bias towards lower ratings, as depicted in \figref{fig:dataset_labels_hist}.
Thus, the model might learn to assign lower ratings in average.

\subsubsection{Ablation study}

In Tab.~\ref{tab:results_corr}, we report the performance of three variants of the model, each with one improvement removed.
This allows us to identify the gain provided by each of them.
First, we observe the influence of the temporal resolution of the features at 40~ms instead of 400~ms as in eMoBi-Q.
This change raises the correlation from 0.85 to 0.90 across the whole test set.
We interpret this improvement as the 40~ms resolution being closer to that of coding artifacts, allowing for better identification.
Second, adding of the diffuseness feature (which is the only one applied on the ambisonic representation) slightly improves correlation coefficients.
Note that the version of the model without the diffuseness feature allows for the evaluation of any binaural signal.
Third, adding a pre-weighting at the beginning of the neural network also improves the correlation criteria, at the cost of a slight increase in RMSE.

\section{CONCLUSION}

In this study, we propose QASTAnet, a new audio quality metric for ambisonic or binaural signals.
This method is based on expert feature extraction and a deep neural network modeling the subjective quality assessment process of an average listener.

The proposed metric is compared to two baseline methods: Ambiqual and eMoBi-Q.
The evaluation focuses on the degradations introduced by codecs on a wide range of content types (speech, music, ambiance, anechoic, reverberated).
The prediction performance of QASTAnet surpasses that of existing metrics, particularly on signals containing realistic spatial reverberation.
Our study suggests that in the absence of an expert model for subjective quality assessment, this function can be approximated using a supervised-trained neural network.
This can be achieved with relatively limited data and noisy labels.
To prevent overfitting, the neural network must limit the number of trainable parameters.
We provide a fully differentiable open-source implementation in Python/Pytorch\footnote{\url{https://github.com/Orange-OpenSource/QASTAnet}}.
This way, the metric can be used as a training objective for the development of spatial audio processing.
Future research may extend the evaluation to study the generalization of QASTAnet's performance to other degradations, \eg, ambisonic order truncation, other codecs, packet loss concealment.
It would also be useful to generalize the metric to other spatial formats, \eg, 5.1, 7.1.4, and to take into account other psychoacoustics effects, \eg, temporal and frequency masking.



\end{document}